\documentstyle[aps,subfig]{revtex}
\input epsf
\begin{document}

\thispagestyle{empty}
\begin{center}
\vspace*{1cm}
{\bf \Large
 Partially  Asymmetric Exclusion Process \\
with Open Boundaries
}\\[10mm]
{\Large {\sc
Sven Sandow
} }\\[9mm]

\begin{minipage}[t]{15cm}
\begin{center}
{\small\sl

  Department of Physics of Complex Systems,\\
             The Weizmann Institute of Science, \\
  Rehovot 76100,
  Israel}

\end{center}
\end{minipage}
\end{center}

{\small Exclusive diffusion  on a one-dimensional lattice is studied. In the
model particles hop stochastically into both directions with different rates.
At the ends of the lattice particles are injected and removed. The exact
stationary probability measure is represented in form of a matrix product as a
generalization of the solution given by Derrida et al \cite{dehp} for the
fully asymmetric process. The phase diagram of the current on the infinite
lattice is obtained. Analytic expressions for the current in the different
phases are derived. The model is equivalent to a $XXZ$-Heisenberg chain with a
certain type of boundary terms the ground state of which corresponds to
the stationary solution of the master equation. }\\[0.2cm]



\section{Introduction} The one-dimensional stochastic exclusion process is of
interest for several reasons. Besides being the simplest example for diffusion
of interacting particles \cite{lig},\cite{spi} it is closely related to various
other phenomena such as interface growth \cite{ks},\cite{km}, the dynamics of
shocks \cite{k},\cite{djls} or of directed polymers  \cite{ks} as well as to
freeway traffic \cite{scsc}-\cite{nag}. Furthermore it can be mapped on vertex
models \cite{kdn} or  quantum spin chains \cite{grs}-\cite{adhr}.  Although the
models are fairly simple only few exact results are known
\cite{dehp}-\cite{djls}, \cite{lps}-\cite{ss}.

Of particular interest is the asymmetric exclusion process with open
boundaries as an example for a driven diffusive system coupled to its
environment. For the fully asymmetric process where particles hop only into
one direction  the exact stationary state is known . The problem was solved by
Derrida, Domany and Mukamel \cite{ddm} for special choices of the system
parameters and more generally by Sch\"utz, Domany \cite{sd} and by Derrida et
al \cite{dehp}. Several phase transitions were found for this model.  The
partially asymmetric process where particles are allowed to hop into both
directions, but with different rates, is a natural generalization of the fully
asymmetric process. Some exact results are known for periodic boundary
conditions \cite{gs} as well as for a closed chain \cite{ss,adhr}. In
 both cases
characteristic time scales can be determined. In case of open boundaries an
algebraic representation of the stationary solution was proposed in
\cite{dehp} for particular choices of the input and output rates.

The aim of  this paper is to study the partially asymmetric process with open
boundaries.  The method used here is the algebraic Ansatz introduced in
\cite{dehp}. Our main result is the stationary solution of the master equation.
{}From this expressions for   the current on a large lattice are deduced (see
eq.
(\ref{C12})-(\ref{C14}) ). Like for the fully asymmetric process three phases
are encountered (see the figure in section IV) assumed that the direction of
the drift is fixed. It turns out that these phases reduce to the ones of  the
fully asymmetric process in the appropriate limit.

The paper is organized as follows. In the next section the model is defined.
Then, in section III the stationary solution is introduced and two
representations are discussed. The stationary current on a large lattice and
its phase diagram are  studied in section IV.   Section V concludes with some
remarks on the density profile,  mean field results, the algebraic
structure of the solution and on the ground state of the corresponding $XXZ$-
chain.

\section{The model}
\renewcommand{\theequation}{\arabic{section}.\arabic{equation}}

We consider a one-dimensional lattice of length $L$. Each lattice site can be
occupied by one particle or can be empty. Hence, the state of the system is
defined by a set of occupation numbers ${\tau_1,...,\tau_L}$ while $\tau_i=1$
($\tau_i=0$) means site $i$ is occupied (free). The particles are assumed to
move stochastically on the lattice. With rate $p$ they hop to their right if
their nearest
 neighbour site on the right is empty, and with a rate $q$ the hop to the left
if their left neighbour site is empty. Particles are injected at the left
(right) boundary with a rate $\alpha$ ($\delta$) and removed on the left
(right)
with a rate $\gamma$ ($\beta$). The dynamics is supposed to be sequential,
i.e.,
only one particle can hop at a time. Thus if the system has the configuration
${\tau_1(t),...,\tau_L(t)}$ at time $t$ it will change to:\\[0.5cm]
for $1<i<L$:
\begin {eqnarray} \label{A1}
\tau_i(t+dt)&=&1 \;\;\;\mbox{with probability}\;\;
 x_i=\tau_{i}(t)+[ \{p\tau_{i-1}(t)+q\tau_{i+1}(t)\}\{1-\tau_i(t)\}\nonumber\\
&&-p\tau_{i}(t)\{1-\tau_{i+1}(t)\}-q\tau_{i}(t)\{1-\tau_{i-1}(t)\}]dt\\
\tau_i(t+dt)&=&0 \;\;\;\mbox{with probability}\;\;\;1-x_i\nonumber
\end {eqnarray}
for $i=1$:
\begin {eqnarray} \label{A2}
\tau_1(t+dt)&=&1 \;\;\;\mbox{with probability}\;\;
x_1=\tau_{1}(t)+[\{\alpha+q\tau_{2}(t)\}\{1-\tau_{1}(t)\}\nonumber\\
&&-\tau_{1}(t)\{p(1-\tau_{2}(t))\}+\gamma\}]dt\\
\tau_1(t+dt)&=&0 \;\;\;\mbox{with probability}\;\;\;1-x_1\nonumber
\end{eqnarray}
and for $i=L$:
\begin{eqnarray} \label{A3}
\tau_L(t+dt)&=&1 \;\;\;\mbox{with probability} \;\;
x_L=\tau_{L}(t)+[\{\delta+p\tau_{L-1}(t)\}\{1-\tau_{L}(t)\}\nonumber\\
&&-\tau_{L}(t)\{q(1-\tau_{L-1}(t))+\beta\}]dt\\
\tau_L(t+dt)&=&0 \;\;\;\mbox{with probability}\;\;\;1-x_L\nonumber
\;\; .
\end{eqnarray}
This defines a master equation for the probability distribution which we may
write as
\begin{eqnarray} \label{A4}
\partial_t P(\tau_1,...,\tau_L,t)=H P(\tau_1,...,\tau_L,t) \\
\label{A5}\mbox{with}\;\;\;
H=h'_1+\sum_{i=1}^{L-1} h_{i,i+1}+h'_L
\;\;. \end{eqnarray}
where the operator $h'_1$ ($h'_L$) describes the change of the probability by
 means of particle input or output at the left (right) boundary and
 $ h_{i,i+1}$
gives the impact of the jumps in the bulk. The excplicit form of the operators
depends on the representation we choose.

The model exhibits two symmetries. It is invariant under the following
exchanges:
\begin{eqnarray}\label{A6}
&\tau_i \leftrightarrow \tau_{L+1-i}\;\;\; \mbox{and}\;\;\; p \leftrightarrow q
\;\;\;\mbox{and}\;\;\;  \alpha\leftrightarrow \delta \;\;\;\mbox{and}\;\;\;
\beta\leftrightarrow \gamma\\
\label{A7} \mbox{or}\;\;\;&
\tau_i \leftrightarrow 1-\tau_{i} \;\;\;\;\;\;\mbox{and}\;\;\; p
\leftrightarrow q
\;\;\;\mbox{and} \;\;\; \alpha\leftrightarrow \gamma \;\;\;\mbox{and}\;\;\;
\beta\leftrightarrow \delta \;\;.
\end{eqnarray}
This enables us to restrict our study to the case $p< q$ while results for
the other part of the parameter space are obtained exploiting one of the the
above symmetries. Furthermore we may restrict to $p+q=1$. Any other value
amounts to a rescaling of time only.

In the next section we are going to derive a stationary solution of the master
equation (\ref{A4}).\\

Let us put  some remarks on
 the relation of the  stochastic model to a quantum spin chain here. We map the
 master equation
 on a imaginary time Schr\"odinger equation (see e.g. \cite{st},\cite{ss1},
\cite{ss}).
In a basis defined by
the vectors $|\tau_1,\tau_2,...,\tau_L \rangle$ the Hamiltonian reads then:
\begin{eqnarray}
H & = & - \sqrt{pq}
 \sum_{j=1}^{L-1}\left[\overline{q}^{-1}
(c^{\dagger}_{j}c_{j+1} - (1-n_j)n_{j+1})
+\overline{q}( c^{\dagger}_{j+1}c_{j} - n_j(1-n_{j+1}) ) \right] \nonumber\\
\label{A5a}
&&-\alpha(c^{\dagger}_1-1+n_1)-\gamma (c_1-n_1)
-\delta(c^{\dagger}_L-1+n_L)-\beta (c_L-n_L)\\
\label{A5b}
&& \mbox{with}\;\;\;\overline{q}=\sqrt{\frac{p}{q}}
\;\;.
\end{eqnarray}
Here the operators $c^{\dagger}_{j}$ and $c_{j}$ create and annihilate
 particles at site $j$. They obey spin commutation relations and may be
 written in terms of Pauli matrices:
$c_j = (\sigma^x_j +i \sigma^y_j)/2$ , $c^{\dagger}_j = (\sigma^x_j
- i \sigma^y_j)/2$ and $n_j=(1-\sigma^z_j)/2$. Using the operator \cite{hs}
\begin{equation}
\label{A5c}
V=\overline{q}^{-\sum_{j=1}^L j n_j}
\end{equation}
the Hamiltonian (\ref{A5a}) can
be transformed into
\begin{eqnarray}\label{A5d}
H' & = &VHV^{-1} \\
&=&- \frac{1}{2}\sqrt{pq}
\sum_{j=1}^{L-1}\left[
\sigma^x_j \sigma^x_{j+1}+\sigma^y_j \sigma^y_{j+1}
+\frac{1}{2}(\overline{q}+\overline{q}^{-1}) \sigma^z_j \sigma^z_{j+1}
-\frac{1}{2}(\overline{q}+\overline{q}^{-1}) \right]
\nonumber\\
\label{A5e}
&&-A^+_1 \sigma^x_1-i A^-_1 \sigma^y_1-B_1 \sigma^z_1
-A^+_L \sigma^x_L-i A^-_L\sigma^y_L-B_L \sigma^z_L+
\frac{1}{2}(\alpha+\beta+\gamma+\delta)\\\label{A5f}
&&\mbox{with}\;\;A^{\pm}_1=\frac{1}{2}(\gamma \overline{q} \pm
\alpha \overline{q}^{-1})\;\;;
\;\;B_1=\frac{1}{2}(\gamma-\alpha)+\frac{1}{4}(\overline{q}-\overline{q}^{-1})\\
\label{A5g}
&&\mbox{and}\;\;A^{\pm}_L=\frac{1}{2}(\beta \overline{q}^L \pm
 \delta \overline{q}^{-L} )\;\;;
\;\;B_L=\frac{1}{2}(\beta-\delta)-\frac{1}{4}(\overline{q}-\overline{q}^{-1})
\;\;\end{eqnarray}
which is a spin-$1/2$ $XXZ$-Hamiltionian with a certain class of boundary
 terms.

\section{The stationary solution}
\setcounter{equation}{0}

An algebraic Ansatz for the stationary measure of the fully asymmetric process
was proposed in \cite{dehp}. Proceeding in the same way we assume
the system to have a stationary probability distribution which can be written
 as
\begin{eqnarray} \label{B1}
P_L(\tau_1,...,\tau_L)&=&\langle 0| \prod_{i=1}^L (\tau_iD+(1-\tau_i)A)|0
\rangle
Z_L^{-1}\\
\label{B2}\mbox{where}\;\;\; Z_L&=&\langle 0|C^L |0 \rangle\;\;;\;\;C=D+A
\end{eqnarray}
with some matrices $D$ and $A$ and vectors $\langle 0|$ and $|0 \rangle $.
An algebra which was shown to give a stationary probability distribution for a
lattice with up to three sites in \cite{dehp} is the following one:
\begin{eqnarray} \label{B3}
p DA-q AD&=&D+A\\
\label{B4}(\beta D-\delta A)|0 \rangle& =&|0 \rangle \\
\label{B5}\langle 0|(\alpha A-\gamma D) & =& \langle 0|
\;\;.\end{eqnarray}
This algebra defines a stationary solution for a lattice of any number of sites
which can be shown as follows.
Consider the action of the operator $h_{i,i+1}$ from (\ref{A5}) on the
probability of a certain state on a lattice of $L+1$ sites:
\begin{eqnarray} \label{B6}
h_{i,i+1}P_{L+1}(...\tau_{i-1}10\tau_{i+2}...)=-pP_{L+1}(...\tau_{i-1}10\tau_{i+
2}...)+qP_{L+1}(...\tau_{i-1}01\tau_{i+2}...)\nonumber
\;\;. \end{eqnarray}
Using (\ref{B3}) results in
\begin{equation} \label{B7}
h_{i,i+1}P_{L+1}(..\tau_{i-1}10\tau_{i+2}...)=
-k\{P_{L}(..\tau_{i-1}0\tau_{i+2}...)+P_{L}(..\tau_{i-1}1\tau_{i+2}...)\}
\end{equation}
where $k=Z_L/Z_{L+1}$ and similarily
\begin{equation} \label{B8}
h_{i,i+1}P_{L+1}(..\tau_{i-1}01\tau_{i+2}...)=
k\{P_{L}(..\tau_{i-1}0\tau_{i+2}...)+P_{L}(..\tau_{i-1}1\tau_{i+2}...)\}
\;\;.\end{equation}
Obviously
\begin{eqnarray} \label{B9}
h_{i,i+1}P_{L+1}(...\tau_{i-1}00\tau_{i+2}...)&=&0\\
\label{B10}\mbox{and} \;\;\;h_{i,i+1}P_{L+1}(...\tau_{i-1}11\tau_{i+2}...)&=&0
\;\;\;.\end{eqnarray}
Using (\ref{B4}) and (\ref{B5}) we find:
\begin{eqnarray} \label{B11}
h'_1 P_{L+1}(1\tau_{2}...)&=&kP_{L}(\tau_{2}...) \\
\label{B12}
h'_1 P_{L+1}(0\tau_{2}...)&=&-kP_{L}(\tau_{2}...) \\
\label{B13}
h'_L P_{L+1}(...\tau_{L}1)&=&-kP_{L}(...\tau_{L}) \\
\label{B13a}
h'_L P_{L+1}(...\tau_{L}0)&=&kP_{L}(...\tau_{L})
\;\;.\end{eqnarray}
Eq. (\ref{B7})-(\ref{B13a}) explicitly describe the action of the local parts
of the Hamiltonian.  Applying the full Hamiltonian $H=h'_1+\sum_{i=1}^L
h_{i,i+1}+h'_L$ to the probability of any configuration
 on a lattice of length $L+1$
all occurring terms add up to zero. Consequently the algebra
(\ref{B3})-(\ref{B5}) defines a stationary probability distribution.

Let us remark that for periodic boundary conditions the stationary solution
can be written as
 $ P_L(\tau_1,...,\tau_L) \propto
\mbox{{\large  Tr }} \prod_{i=1}^L (\tau_iD+(1-\tau_i)A) $ with the  algebra
(\ref{B3}). This can be proved the same way as above.\\

The algebra  (\ref{B3})-(\ref{B5}) in the above given  form does not allow for
explicit calculation of probabilities. However, the following relations,
equivalent to eqs. (\ref{B3})-(\ref{B5}), define rules to compute
$P_L(\tau_1...\tau_L)$ :
\begin{eqnarray} \label{B14}
{}^{(1)}\langle k|&=&\langle 0|D^k\\
\label{B15}
{}^{(1)}\langle k|A&=&\sum_{i=0}^{k+1} a_{ki}\;{}^{(1)}\langle i|\\
\label{B16}
{}^{(1)}\langle k|0\rangle&=&s_k\;\;\forall k=0,1,2,...
\end{eqnarray}
with
\begin{eqnarray} \label{B17}
a_{k0}&=&p^{-k} \alpha^{-1}\;\;\;\forall k\ge 0\\
a_{ki}&=&\alpha^{-1}
\renewcommand{\arraystretch}{0.8}
\mbox{$\left(\begin{array}{@{}c@{}}{\scriptstyle
k}\\{\scriptstyle i}\end{array}\right)$}\renewcommand{\arraystretch}{1}
p^{-k}q^i\{\gamma q^{-1} \frac{i}{k-i+1}+1\}
\nonumber\\\label{B18}
&&+\sum_{\nu=0}^{i-1}
\renewcommand{\arraystretch}{0.8}
\mbox{$\left(\begin{array}{@{}c@{}}{\scriptstyle
k-i+\nu}\\{\scriptstyle i-1
}\end{array}\right)$}\renewcommand{\arraystretch}{1}
p^{-k+i-\nu-1}q^{\nu}\;\;\;
\forall k\ge 0\;;\; 0<i<k+1\\
\label{B19}
a_{kk+1}&=&p^{-k}q^k \alpha^{-1} \gamma\;\;\;\forall k\ge 0
\end{eqnarray}
and
\begin{equation} \label{B20}
\beta s_{k+1}=\delta
\sum_{i=0}^{k+1} a_{ki}s_i+s_k
\;\;.\end{equation}
Here we assumed $\alpha>0$.\\
Eq. (\ref{B14}) is the definition of the vectors ${}^{(1)}\langle k|$. Eq.
(\ref{B15}) and  (\ref{B17})-(\ref{B19}) are consequences of eq. (\ref{B3}) for
the product $DA$   and eq. (\ref{B5}). Eqs. (\ref{B16}) and (\ref{B20}) follow
from (\ref{B4}). A detailed proof of the equivalence between the algebra
(\ref{B3})-(\ref{B5}) and the rules (\ref{B14})-(\ref{B20}) is given in
appendix A.

Note that in the case of totally asymmetric diffusion ($p=1;q=\gamma=\delta=0$)
studied in \cite{dehp,sd} eqs. (\ref{B14})-(\ref{B20}) simplify to
\begin{eqnarray} \label{B21}
{}^{(1)}\langle k|&=&\langle 0|D^k\\
\label{B22}
{}^{(1)}\langle k|A&=&\sum_{i=1}^k \;{}^{(1)}\langle i|+\alpha^{-1}\langle 0|\\
\label{B23}
{}^{(1)}\langle k|0\rangle&=&\beta^{-k}\;\;\forall k=0,1,2,...
\;\;.\end{eqnarray}

Another similar representation of the algebra (\ref{B3})-(\ref{B5}) can be
found
starting from $|k\rangle=A^k|0\rangle$ and proceeding in a way as above.

The stationary probability distribution is given by eq. (\ref{B1}). In order to
find the probability of a certain configuration we have to calculate
expressions
like $ \langle 0|DADDDAA...|0\rangle$ which reduce to linear combinations of
the scalar products $s_k$ after application of rules (\ref{B14})-(\ref{B19}).
Fixing $s_0$, say  to $s_0=1$, the latter quantities can be computed from the
recursion (\ref{B20}). Hence  eqs. (\ref{B1}),(\ref{B2}) combined with eqs.
(\ref{B14})-(\ref{B20}) give a representation of the stationary probability
distribution and enables us to calculate the probability of any configuration
as well as any kind of averaged quantity. They may be used for numerical
calculations. The dimension of the representation is $L+1$.\\

We introduce another representation of the algebra (\ref{B3})-(\ref{B5}) here
which is more convenient for the large lattice approximations done in the next
section. Let us start with defining operators $F$ and $F^{\dagger}$ by
\begin{eqnarray} \label{B24}
D&=&\frac{1}{p-q} (F+1)\\
\label{B25}
A&=&\frac{1}{p-q} (F^{\dagger}+1)
\;\;.\end{eqnarray}
The relation (\ref{B3}) reads in terms of $F$ and $F^{\dagger}\;$
\begin{equation} \label{B26}
F^{\dagger}F-\overline{q}^2 FF^{\dagger}=1-\overline{q}^2
\;\;;\;\;(\;\overline{q}=\sqrt{\frac{p}{q}}>1\;)
\;\;.\end{equation}
Operators commuting as above are known to be related to  creation and
annihilation operators
of a q-deformed harmonic oscillator \cite{mac,bie,kul}. The latter ones may
be defined as
$a^{\dagger}=(\overline q-\overline q^{-1})^{-1/2}
  \overline q^{N/2} F^{\dagger}$ and
$a=(\overline q-\overline q^{-1})^{-1/2}
 F \overline q^{N/2}$ where $N$ is a particle number operator \cite{mac}.

We choose a representation of (\ref{B26}) as \cite{mac}:
 \begin{eqnarray}
\label{B27}
 F^{\dagger} |k \rangle^{(2)}&=& \{k+1\}_{\overline q}^{1/2}
|k+1\rangle^{(2)}\\
 \label{B28}
F |k \rangle^{(2)}&=& \; \{k\}_{\overline q}^{1/2}
|k-1 \rangle^{(2)}\\
\label{B29}
 \mbox{with}&&\;\{x\}_{\overline q}=1-{\overline q}^{-2x}\;\;
\mbox{for any}\;\;x>0
\end{eqnarray}
where the $|k \rangle^{(2)}$ with  $k=0,1,2,...$ form an orthogonal basis in
an infinite dimensional Hilbert space.
 $F^{\dagger}$ is adjoint to $F$. Eqs.
(\ref{B24}),(\ref{B25}) and (\ref{B27})-(\ref{B29}) specify the action of the
 operators $D$ and $A$ in this basis. Note that up to factor the q-oscillator
 operators $a^{\dagger}$ and $a$ act like  $F^{\dagger}$ and $F$ here. The
consistence of the representation with eqs. (\ref{B26}) is obvious.

 For the computation of probabilities according to (\ref{B1}) we need
furthermore  to project the vectors $\langle0|$ and $|0 \rangle$ on the basis
vectors which has to be done in a way that  the relations (\ref{B4}) and
(\ref{B5}) representing the boundary conditions are ensured.  Let us write
\begin{eqnarray} \label{B30}
|0 \rangle& =&\sum_{i=0}^{\infty}r_i  \;|i \rangle^{(2)} \\
\label{B31}
\langle 0|& =&\sum_{i=0}^{\infty}l_i \;\; {}^{(2)}\langle i|
\;\;.
\end{eqnarray}
Eq. (\ref{B4}) reads in terms of $F^\dagger$ and $F$:
\begin{equation} \label{B32}
[\beta F-\delta F^\dagger+\beta-\delta-p+q]|0 \rangle=0
\;\;\end{equation}
which results after using eqs. (\ref{B30}) as well as (\ref{B27}) and
(\ref{B28}) in
\begin{eqnarray} \label{B33}
0&=&\beta \{k+1\}_{\overline q}^{1/2} r_{k+1}
+(\beta-\delta-p+q)r_k-\delta \{k\}_{\overline q}^{1/2} r_{k-1}\\
\label{B34}
\mbox{where}&&r_0=1\;\;\mbox{and}\;\;r_{-1}=0
\;\;.\end{eqnarray}
{}From eq. (\ref{B5}) we find for the left vector in the same way:
\begin{eqnarray} \label{B35}
0&=&\alpha \{k+1\}_{\overline q}^{1/2} l_{k+1}
+(\alpha-\gamma-p+q)l_k-\gamma \{k\}_{\overline q}^{1/2} l_{k-1}\\
\label{B36}
\mbox{where}&&l_0=1\;\;\mbox{and}\;\;l_{-1}=0
\;\;.\end{eqnarray}
Eq. (\ref{B35}) can be obtained from eq. (\ref{B33}) by replacing $r_k$ by
$l_k$, $\beta$ by $\alpha$ and $\delta$ by $\gamma$.   Eqs.
(\ref{B30}) and  (\ref{B31}) give a
representation of the boundary vectors in the above defined basis. Together
with eq. (\ref{B24}), (\ref{B25}) and (\ref{B27})-(\ref{B29}) which specify
the action of $D$ and $A$  we may calculate any probability supposed we have
computed the coefficient $r_k$ and $l_k$ by means of the recursions
(\ref{B33}) and (\ref{B35}). The procedure is not very convenient for
  numerical examples since we have to deal with an infinite dimensional
representation.
However large lattice approximations are to be seen easier here
than in the first representation. The reason is that the impact of the left
boundary, of the right boundary and of the bulk enter in the left vector, in
 the right vector and in the matrices, respectively.\\

The physical quantity we are to analyze in detail is the current $j$ which in
the stationary state is independent on the position. It can be written
in a convenient way as:
\begin{equation} \label{B37}
j=\langle 0|C^{L-2}(pDA-qAD) C|0\rangle Z_L^{-1}=Z_{L-1} Z_L^{-1}
\;\;\end{equation}
because of eqs. (\ref{B1})-(\ref{B3}). Therefore we have to calculate
$Z_L=\langle 0|C^L|0\rangle$. Let us define the coefficients
$c_{ik}^L=\;\;{}^{(2)}\langle i|C^{L}|k\rangle^{(2)}$
as the matrix elements of the operator $C^L$.
Applying $C=(F^{\dagger}+F+2)/(p-q)$ from the left to
\begin{equation} \label{B38}
C^L |k\rangle^{(2)}= \sum_{i=0}^{\infty}
c_{ik}^L|i\rangle^{(2)}
\;\;.\end{equation}
 and using the rules
(\ref{B27}), (\ref{B28}) we get a recursion relation:
\begin{eqnarray} \label{B39}
c_{ik}^{L+1}&=& \frac{1}{p-q}
[\;\{i\}_{\overline q}^{1/2}c_{i-1k}^{L}
+2c_{ik}^{L}+\{i+1\}_{\overline q}^{1/2}c_{i+1k}^{L}\;] \\
\label{B40}
\mbox{with} \;\;c_{ik}^0&=&\delta_{i,k}\;\;\mbox{and}\;\;
c_{ik}^L=0\;\;\mbox{for}\;\;i>k+L\;\;\mbox{or}\;\;i<max\{k-L,0\}
\;\;.\end{eqnarray}
Using the the decompositions of the boundary vectors (\ref{B30}) and
(\ref{B31}) as well as
eq. (\ref{B38}) we find the following expression for
 $Z_L=\langle 0| C^L |0\rangle$:
\begin{equation} \label{B41}
Z_L=\sum_{k,i=0}^{\infty}
l_i c_{ik}^{L}r_k
\;\;.
\end{equation}
After solving the recursions for $l_k$, $r_i$ and $c_{ki}^{L}$ the above
equation allows for the calculation of $Z_L$ and hence of the current. In the
next section we are going to do this for a large lattice.

\section{The stationary current for a large lattice}
\setcounter{equation}{0}

The following approximation is derived in appendix B:
\begin{eqnarray} \label{C1}
r_k \propto\;\;\left\{
\begin{array}{ll}
(\kappa_+(\beta,\delta))^k \;\;\mbox{for}\;\;\kappa_+(\beta,\delta)>1
 \;\;\mbox{and}\;\;k \gg 1\\
(\kappa_-(\beta,\delta))^k \;\;\mbox{for}\;\;\kappa_+(\beta,\delta)<1
\;\;\mbox{and}\;\;k \gg 1
\end{array}
\right.
\end{eqnarray}
with
\begin{equation} \label{C2}
\kappa_{\pm}(\beta,\delta)=\frac{1}{2 \beta}[-\beta+\delta+p-q \pm
\sqrt{(-\beta+\delta+p-q)^2+4 \beta \delta}\;]
\;\;.\end{equation}
For all choices of parameters $|\kappa_{-}(\beta,\delta)|<1$. That means for
 $\kappa_{+}(\beta,\delta)<1$ the coefficient $r_k$ decreases exponentially
with
$k$ whereas it increases exponentially for $\kappa_{+}(\beta,\delta)>1$.
Note that the condition $\kappa(\beta,\delta)>1$ is equivalent to
$\beta-\delta<\frac{p-q}{2}$.\\

Similarly we find for the coefficients $l_k$:
\begin{eqnarray} \label{C3}
l_k \propto\;\;\left\{
\begin{array}{ll}
(\kappa_+(\alpha,\gamma))^k \;\;\mbox{for}\;\; \kappa_+(\alpha,\gamma)>1
\;\;\mbox{and}\;\;k\gg 1 \\
(\kappa_-(\alpha,\gamma))^k \;\;\mbox{for}\;\; \kappa_+(\alpha,\gamma)<1
\;\;\mbox{and}\;\;k\gg 1 \\
\end{array}
\right.
\end{eqnarray}
with
\begin{equation} \label{C4}
\kappa_{\pm}(\alpha,\gamma)=\frac{1}{2 \alpha}[-\alpha+\gamma+p-q \pm
\sqrt{(-\alpha+\gamma+p-q)^2+4 \alpha \gamma}\;]
\;\;
\end{equation}
and $|\kappa_{-}(\alpha,\gamma)|<1$. That means for
 $\alpha-\gamma>\frac{p-q}{2}$ where
 $\kappa_{+}(\alpha,\gamma)<1$ the coefficient $r_k$ decreases exponentially
 with $k$ whereas it increases exponentially for $\kappa_{+}(\alpha,\gamma)>1$.

The coefficients $c_{ik}^L$ are found to obey a simple relation:
\begin{equation} \label{C5}
c_{ik}^L/c_{ik}^{L-1} \approx (\frac{4}{p-q}) \;\;\mbox{for}\;\;
k \ll L\;\;,\;\;i \ll L\;\;\mbox{and}\;\;L \gg 1
\;\;.\end{equation}
This can be shown analytically for $\overline q \rightarrow \infty$, where a
recursion of the same type as for the $c_{ik}^L$ was solved  in \cite{dehp}. We
did not succeed in deriving eq. (\ref{C5}) for general $ \overline q$, but the
recursion (\ref{B39}) can be solved numerically. For all choices of  $
\overline q$  the above relation turns out to be correct. This result is
plausible since after removing the factor  $1/(p-q)$ the recursion is  similar
to that for the simpler special case. Only additional factors $\{k\}_{\overline
q}^{1/2}$ occur which are approximately $1$ for almost all $k$.

We use this approximations for the calculation of $Z_L$ by means of eq.
 (\ref{B41}). Different cases have to be distinguished.\\

{\em The case $\kappa_+(\beta,\delta)<1$ , $\kappa_+(\alpha,\gamma)<1$.} Here
both of the coefficients $r_i$, $l_i$ fall exponentially with $i$.
 Hence the main
contribution to $Z_L$ is given by the terms with small $i,k$ in the sum. For
small $i,k$ the approximation (\ref{C5}) for the $c_{ik}^L$ can be used. The
region of validity of (\ref{C5}) increases with $L$ while the descent of
$r_i$ and $l_i$ is independent of $L$.
Therefore  we find for $L \gg 1\;\;$ $Z_L /Z_{L-1} \approx \frac{4}{p-q}$ and
for the current:
\begin{equation} \label{C11}
j \approx \frac{p-q}{4}
\;\;.\end{equation}\\

{\em The case $\kappa_+(\beta,\delta)>\kappa_+(\alpha,\gamma)$ ,
$\kappa_+(\beta,\delta)>1$.} Here the coefficient $r_k$ increases
exponentially with $k$. $l_i$ may or may not increase with $i$. In any
case $l_i$ cannot increase faster than $r_i$.
Suppose now $l_i$ increases as well. Then the sum (\ref{B41}) does not exist
 since it extends  over terms like
$[\kappa_+(\beta,\delta) \kappa_+(\alpha,\gamma)]^k$. In order to avoid this
divergence we redefine $\langle 0|$ as
 $\langle 0| = \lim_{N \rightarrow \infty}\{\;
\sum_{i=0}^{N}l_i \;\; {}^{(2)}\langle i|\;\;
[\sum_{j=0}^{N} \lambda^j]^{-1}\;\}$ with some real $\lambda$.
 Chosing $\lambda>\kappa_+(\beta,\delta)
 \kappa_+(\alpha,\gamma)$ ensures convergence of the sum for $Z_L$. Consider
 now the  product $\langle 0|C^{L-1}F|0\rangle$. Since
$ F|0\rangle=\sum_{k=0}^{\infty}\{k+1\}_{\underline q}^{1/2}r_{k+1} |k\rangle$
we may write
\begin{eqnarray}
\langle 0|C^{L-1}F|0\rangle
&=&\lim_{N \rightarrow \infty}\{\;\sum_{i=0}^N \sum_{k=0}^{\infty}
 \{k+1\}_{\underline q}^{1/2}\;
l_i c_{ik}^{L-1}r_{k+1}\;
[\sum_{j=0}^{N} \lambda^{j}]^{-1} \;\}
\nonumber\\
\label{C6}
&\approx&\kappa_+(\beta,\delta) Z_{L-1}
\;\;\end{eqnarray}
where we have used the fact that the main contribution to the $k$-sum
for any $i$ stems from
large $k$ terms. Small $k$ terms with large $i$ contribute less to the whole
sum than large $k$ terms since $r_k$ increases faster than $l_i$.
 And for
$k \gg1 $ we can approximate
$r_k\propto(\kappa_+(\beta,\delta))^k$ and
$\{k+1\}_{\overline q}^{1/2}\approx 1$ .
In case $l_i$ does not increase with $i$ the same argument applies.

 Writing now  $C$ as
    $C=\delta^{-1}[\;-(\beta D-\delta A)+\frac{1}{p-q}(\beta+\delta)
+\frac{1}{p-q}(\beta+\delta)F \;]$ we get
\begin{eqnarray}
Z_L&=&\delta^{-1}\langle 0|C^{L-1}\{
-(\beta D-\delta A)+\frac{1}{p-q}(\beta+\delta)
+\frac{1}{p-q}(\beta+\delta)F
\}|0 \rangle\nonumber\\
\label{C7}
&=&\delta^{-1}\frac{1}{p-q}
\{(\beta+\delta-p+q)
Z_{L-1}+(\beta+\delta)\langle 0|C^{L-1}F |0 \rangle \}
\end{eqnarray}
Inserting eq. (\ref{C6}) into eq. (\ref{C7}) results in
\begin{equation}  \label{C8}
Z_L \approx Z_{L-1} \delta^{-1}\frac{1}{p-q}
\{(\beta+\delta-p+q)
+(\beta+\delta) \kappa_+(\beta,\gamma) \}
\;\;.\end{equation}
Finally, using eq. (\ref{C2}) for $\kappa_+(\beta,\gamma)$ we find for
the current
$j \approx Z_{L-1}/Z_L$:
\begin{equation} \label{C9}
j \approx \frac{1}{2(p-q)}\;\{ \;(\beta-\delta)(p-q)-(\beta+\delta)^2+
(\beta+\delta)
\sqrt{(\beta-\delta-p+q)^2+4 \beta \delta}\; \}
\;\;.\end{equation}

{\em The case $\kappa_+(\alpha,\gamma)>\kappa_+(\beta,\delta)$ ,
$\kappa_+(\alpha,\gamma)>1$.} The current for this case can be derived simply
from the above result by applying both symmetry operations (\ref{A6}) and
(\ref{A7}). Since the current does not depend on the position on the lattice
we just have to replace $\beta$ by $\alpha$ and $\delta$ by $\gamma$
in eq. (\ref{C9}):
\begin{equation} \label{C10}
j \approx \frac{1}{2(p-q)}\;\{ \;(\alpha-\gamma)(p-q)-(\alpha+\gamma)^2+
(\alpha+\gamma)
\sqrt{(\alpha-\gamma-p+q)^2+4 \alpha \gamma}\; \}
\;\;.\end{equation}

{\em Summary} Assuming above approximations to be exact for
 $L \rightarrow \infty$ we have found analytic expressions for the current $j$
on an infinite lattice. The dependence of $j$ on the system parameters is
different in three regions of the parameter space. Fig.1 shows the phase
 diagram. The phase separation lines are defined
by means
of the nonlinear expressions $\kappa_+(\beta,\delta)$ and
$\kappa_+(\alpha,\gamma)$ (see eq. (\ref{C2}), (\ref{C4}) ). In the fully
asymmetric limit these
functions reduce to $\kappa_+(\alpha,\gamma)=(1-\alpha)/\alpha$,
$\kappa_+(\beta,\delta)=(1-\beta)/\beta$ for $\alpha,\beta<1$
and the phase diagram is the one described in
\cite{dehp}, \cite{sd}.

\begin{figure}[htbtv]
\begin{center}
\setlength{\epsfysize}{80mm}
\leavevmode
\epsffile{phases.ps}
\vspace{2mm}
\caption{Phase diagram for the current on a large lattice for  $p>q$ in
terms of $\kappa_+(\alpha,\gamma)$ and $\kappa_+(\beta,\delta)$ where
  $\kappa_+(x,y)=\frac{1}{2 x}
[-x+y+p-q + \sqrt{(-x+y+p-q)^2+4 x y}\;]$.
The phases are separated by the  dashed lines.}
\end{center}
\end{figure}

The currents in the three phases are:\\
{\bf Phase A:} $\;\;\kappa_+(\beta,\delta)>\kappa_+(\alpha,\gamma)$ ,
$\kappa_+(\beta,\delta)>1$
\begin{equation} \label{C12}
j=\frac{1}{2(p-q)}\;\{ \;(\beta-\delta)(p-q)-(\beta+\delta)^2+(\beta+\delta)
\sqrt{(\beta-\delta-p+q)^2+4 \beta \delta}\; \}
\end{equation}
Note that the condition $\kappa_+(\beta,\delta)>1$ is fulfilled if
 $\beta-\delta<\frac{p-q}{2}$. In the fully asymmetric limit , i.e., for
$p=1$, $q=\gamma=\delta=0$, the high density phase described in \cite{sd,dehp}
is recovered.\\
{\bf Phase B:}$\;\;\kappa_+(\alpha,\gamma)>\kappa_+(\beta,\delta)$ ,
$\kappa_+(\alpha,\gamma)>1$
\begin{equation} \label{C13}
j=\frac{1}{2(p-q)}\;\{ \;(\alpha-\gamma)(p-q)-(\alpha+\gamma)^2+
(\alpha+\gamma)
\sqrt{(\alpha-\gamma-p+q)^2+4 \alpha \gamma}\; \}
\;\;.\end{equation}
The condition $\kappa_+(\alpha,\gamma)>1$ is fulfilled if
 $\alpha-\gamma<(p-q)/2$. This phase corresponds to the low density phase in
the
fully asymmetric limit.\\
{\bf Phase C:}$\;\;\kappa_+(\beta,\delta)<1$ , $\kappa_+(\alpha,\gamma)<1$
\begin{equation} \label{C14}
j=\frac{p-q}{4}
\;\;.\end{equation}
In the fully asymmetric limit the maximum current
 phase is recovered.

These phases were found for $p>q$. For $p<q$ similar results are derived
in a trivial manner  using one of the symmetries (\ref{A6}) or (\ref{A7}).

Numerical calculations can be done easiest in the first representation of the
algebra. They show that for typical choices of parameters (not to close to the
phase lines) in the phases A or B   the current has  its value given above
already for $L<50$. In the phase C the convergence is much slower (about
factor 10).

\section{Conclusion}
\setcounter{equation}{0}

Fixing the preferred direction of the diffusion three phases have been found
for which
the current on an infinite lattice obeys different
equations. They correspond to the phases known for the fully
asymmetric process which are  recovered in the appropriate limit. The
transition
lines as well as the current are described by nonlinear functions of the system
parameters. The full phase structure of the process may be richer taking into
account transitions in the behaviour of other quantities such as the
density profile  which
have not been  studied here. We argue that the situation here is similar to the
one observed in the fully asymmetric model \cite{sd}.

A mean field approximation can be applied in a way similar to what was done for
the fully asymmetric process \cite{ddm}. Like there it turns out that the
obtained current and the phase structure is exact.

The density profile is expected to be similar to the one for the fully
asymmetric process. As it can be seen by numerical calculations or by simple
arguments using the first representation discussed in section III the density
profile in phases A and B approaches a constant value from one of the
boundaries. In phase C the density is constant in the bulk while it varies near
both of the boundaries.

The stationary solution exhibits an interesting algebraic structure. The
commutation relations  (\ref{B26}) are related to the ones of creation and
annihilation operators of  a q-deformed oscillator. The latter ones can be use
to construct  the generators of the quantum group $U_q[SU(2)]$  \cite{mac}
which
is the symmetry group of asymmetric diffusion on a closed chain \cite{ss,adhr}.
  On the other hand, the q-oscillator algebra can be obtained from the quantum
group $U_q[SU(2)]$ as a large $j$-limit \cite{kul}.

Since the stochastic model can be mapped on a spin-$1/2$ $XXZ$-chain its
stationary solution corresponds to the ground state of this quantum system.
Precisely speaking, the state
$|\psi_0\rangle=\sum_{\{\tau_i\}}P_L(\tau_1,,...,\tau_L)
|\tau_1,,...,\tau_L\rangle$ with $P_L(\tau_1,,...,\tau_L) $ given in eq.
(\ref{B1}) is the ground state of the Hamiltonian  (\ref{A5a}). Here $\tau_j=0$
means spin up and $\tau_j=1$ means spin down. The ground state of the
Hamiltonian $H'$ defined by eqs. (\ref{A5e})-(\ref{A5g}) is
$V |\psi_0\rangle=\overline{q}^
{\frac{1}{2}\sum_{j=1}^L j \sigma^z_j}|\psi_0\rangle$.

\section*{Acknowledgments}
The author would like to thank D. Mukamel for stimulating discussions and for
a careful reading of the manuscript.  Financial support by the Minerva
 Foundation is gratefully acknowledged.

\appendix
\setcounter{section}{0}
\renewcommand{\theequation}{\Alph{section}.\arabic{equation}}
\section{}
\setcounter{equation}{0}

We proof the equivalence between the algebras  (\ref{B3})-(\ref{B5}) and
(\ref{B14})-(\ref{B20}) here.
Eq. (\ref{B14}) is the definition of the vectors ${}^{(1)}\langle 0|$.
 Eq. (\ref{B15})
reduces for $k=0$ to eq. (\ref{B5}). To show its equivalence to
(\ref{B3}),(\ref{B5}) for any $k$ we have to apply $A$ to $\langle k|$
iteratively, i.e:
\begin{eqnarray} \label{AA1}
\langle 0|A&=&\alpha^{-1} \langle 0|+\alpha^{-1} \gamma\;{}^{(1)}\langle 1|
\nonumber\\
{}^{(1)}\langle 1|A&=&\langle 0|DA
\nonumber\\
&=&p^{-1}\langle 0|(D+A+qAD)
\nonumber\\
&=&p^{-1}\{q \alpha^{-1}\gamma \;{}^{(1)}\langle 2|+
(1+ \alpha^{-1}\gamma+q\alpha^{-1})\;{}^{(1)}\langle 1|+\alpha^{-1}\langle 0|\}
\nonumber\\
{}^{(1)}\langle 2|A&=&{}^{(1)}\langle 1|DA=...\;\;\;.
\end{eqnarray}
{}From this we see
\begin{equation} \label{AA2}
{}^{(1)}\langle k|A=\sum_{i=0}^{k+1} a_{ki}\;{}^{(1)}\langle i|
\end{equation}
with
\begin{eqnarray} \label{AA3}
a_{k0}&=&p^{-k}\alpha^{-1}\;\;\;\forall k \ge 0\\
a_{ki}&=&\alpha^{-1}
\renewcommand{\arraystretch}{0.8} \mbox{$\left(\begin{array}{@{}c@{}}
{\scriptstyle
k}\\{\scriptstyle i}\end{array}\right)$}
\renewcommand{\arraystretch}{1}
p^{-k}q^i\{\gamma q^{-1} \frac{i}{k-i+1}+1\}
\nonumber\\
\label{AA4}
&&+p^{-k}a'_{ki}\;\;\;
\forall k \ge 0\;;\; 0<i<k+1\\
\label{AA5}
a_{kk+1}&=&p^{-k}q^k \alpha^{-1} \gamma\;\;\;\forall k \ge 0
\;\;.\end{eqnarray}
The coefficients $a'_{ki}$ obey the following recursion relation:
\begin {equation} \label{AA6}
a'_{ki}=p a'_{k-1 i-1}+
\renewcommand{\arraystretch}{0.8} \mbox{$\left(\begin{array}{@{}c@{}}
{\scriptstyle
k-1}\\{\scriptstyle i-1}\end{array}\right)$}
\renewcommand{\arraystretch}{1}
q^{i-1}\;\;\forall 0<i\le k
\end{equation}
which results in
\begin{equation} \label{AA7}
a'_{ki}=\sum_{\nu=0}^{i-1}
\renewcommand{\arraystretch}{0.8} \mbox{$\left(\begin{array}{@{}c@{}}
{\scriptstyle
k-i+\nu}\\{\scriptstyle \nu}\end{array}\right)$}
\renewcommand{\arraystretch}{1}
q^{\nu}p^{i-\nu-1}
\;\;.\end{equation}
Inserting eq. (\ref{AA7}) into (\ref{AA4}) results in the expression
(\ref{B18})
for the coefficients $c_{ki}$ which proofs the equivalence of eq. (\ref{B3})
and (\ref{B5}) to eq. (\ref{B14}),(\ref{B15}),(\ref{B17})-(\ref{B19}).

Multiplying ${}^{(1)}\langle k|$ from the left to eq. (\ref{B4}) and using eq.
(\ref{B16}) gives
\begin{equation}  \label{AA8}
{}^{(1)}\langle k|(\beta D-\delta A)|0\rangle=s_k
\;\;.\end{equation}
Using eqs. (\ref{B14})-(\ref{B15}) we find
\begin{equation} \label{AA9}
\beta s_{k+1}= \delta
\sum_{i=0}^{k+1} a_{ki}s_i+s_k
\end{equation}
which is eq. (\ref{B20}). This completes the proof of the equivalence of the
algebras (\ref{B14})-(\ref{B20}) on one hand and (\ref{B3})-(\ref{B5}) on the
other hand.

\section{}
\setcounter{equation}{0}

In this appendix the large $k$-approximation (\ref{C1}), (\ref{C2}) for the
 components of the right boundary vector $r_k$ is derived. For $k \gg 1$ the
 recursion relation (\ref{B33}) reads
\begin{equation} \label{BB1}
0=\beta  r_{k+1}
+(\beta-\delta-p+q)r_k-\delta r_{k-1}
\;\;.\end{equation}
It is solved by the Ansatz $r_k=b \kappa^k$ which results in
\begin{equation} \label{BB2}
0=\beta  \kappa^2
+(\beta-\delta-p+q) \kappa-\delta
\;\;.\end{equation}
This quadratic equation for $\kappa$ has two solutions:
\begin{equation} \label{BB3}
\kappa_{\pm}(\beta,\delta)=\frac{1}{2 \beta}[-\beta+\delta+p-q \pm
\sqrt{(-\beta+\delta+p-q)^2+4 \beta \delta}\;]
\;\;.\end{equation}
Hence we may approximate for $k\gg1$:
\begin{equation} \label{BB4}
r_k=b_+ (\kappa_{+}(\beta,\delta))^k+b_- (\kappa_{-}(\beta,\delta))^k
\;\;.\end{equation}
{}From eq. (\ref{BB3}) we see that for $\beta-\delta<(p-q)/2$
$\;\;\kappa_{+}(\beta,\delta)>1$ whereas  for all choices of parameters
$\kappa_{-}(\beta,\delta)<1$. And for $\kappa_{+}(\beta,\delta)>1$ the second
decreasing term in eq. (\ref{BB4}) is negligible. On the other hand, if
$\beta-\delta>(p-q)/2$, i.e., $\kappa_{+}(\beta,\delta)<1$, we still find
$|\kappa_{-}(\beta,\delta)|<1$ and
$|\kappa_{-}(\beta,\delta)|>|\kappa_{+}(\beta,\delta)|$. Than the first term in
eq. (\ref{BB4}) becomes negligible in comparison to the second one.
Consequently
\begin{eqnarray} \label{BB5}
 r_k \propto\;\;\left\{ \begin{array}{ll}
(\kappa_+(\beta,\delta))^k \;\;\mbox{for}\;\;\kappa_+(\beta,\delta)>1
\;\;\mbox{and}\;\;k \gg 1\\ (\kappa_-(\beta,\delta))^k
\;\;\mbox{for}\;\;\kappa_+(\beta,\delta)<1 \;\;\mbox{and}\;\;k \gg 1
\end{array} \right.
\end{eqnarray}
This is the large $k$ approximation for the $r_k$.
For the left boundary vector components $l_k$ we find a similar expression
since the recursion for $l_k$ has the same structure as the one for $r_k$. Only
$\beta$ is replaced by $\alpha$ and $\gamma$ is replaced by $\delta$.

\bibliographystyle{unsrt}

\end{document}